\documentclass[aps,pra,superscriptaddress,floatfix,reprint,showpacs,showkeys]{revtex4-1}
\pdfoutput=1
\usepackage{graphicx}
\usepackage{amsmath,amssymb,amstext}
\usepackage{textcomp}

\def\87Rb{$^{87}$Rb}
\def\7Li{$^{7}$Li}
\def\6Li{$^{6}$Li}

\newcommand{\aad}{a_{\textrm{ad}}}

\newcommand{\amax}{a_{\textrm{max}}}

\newcommand{\amaxli}{a_{\textrm{max,Li}}}
\begin{document}

% \preprint{Rev.\ Sci.\ Instr.}

\title{Two-element Zeeman Slower for Rubidium and Lithium}

% \author{Bugs Bunny}
% \affiliation{Looney Tune Studios}
% \author{Roger Rabbit}
% \affiliation{Looney Tune Studios}
% \affiliation{Disney World}
% \author{Mickey Mouse}
% \affiliation{Disney World}

\author{G. Edward Marti}%\email{emarti@berkeley.edu}
\author{Ryan Olf}
\author{Enrico Vogt}
\altaffiliation[Present address: ]{Cavendish Laboratory, University of Cambridge, JJ Thompson Avenue, Cambridge CB3 0HE, United Kingdom}
\author{Anton \"Ottl}
\affiliation{Department of Physics, University of California, Berkeley, California 94720, USA}
\author{Dan M. Stamper-Kurn}
\affiliation{Department of Physics, University of California, Berkeley, California 94720, USA}
\affiliation{Materials Sciences Division, Lawrence Berkeley National Laboratory, Berkeley, California 94720, USA}

\date{\today}

\begin{abstract}
We demonstrate a two-element oven and Zeeman slower that produce simultaneous and overlapped slow beams of rubidium and lithium. The slower uses a three-stage design with a long, low acceleration middle stage for decelerating rubidium situated between two short, high acceleration stages for aggressive deceleration of lithium. This design is appropriate for producing high fluxes of atoms with a large mass ratio in a simple, robust setup.
\end{abstract}



\maketitle

% \tableofcontents

In recent years, experiments that produce quantum gases comprised of multiple atomic elements have enabled a wide range of new investigations.  In some cases, the gaseous mixture is used as a technical means to perform improved studies of a single-element gas, e.g.\ allowing for sympathetic evaporative cooling of an atomic elements with unfavorable collisional properties \cite{Schreck2001,Bloch2001,Modugno2001,Hadzibabic2002} or by allowing one element to act as a collocated detector for a second quantum gas (a thermometer in Ref.~\onlinecite{Catani2009,Spiegelhalder2009,Nascimbene2009} and calorimeter in Ref.~\onlinecite{Catani2009}). In other cases, the gaseous mixture itself is the subject of inquiry.  Novel types of degenerate Bose-Bose \cite{Modugno2002,Catani2008}, Bose-Fermi \cite{Hadzibabic2002,Roati2002,Goldwin2004,Silber2005,Gunter2006,Ospelkaus2006} and Fermi-Fermi \cite{Taglieber2008} mixtures have been produced and used to investigate coherence, superfluidity and disorder. Mixtures of elements with a large mass ratio are of particular interest for studying novel superfluid properties such as spin impurities \cite{Iskin2009}, breached pair superfluids \cite{Liu2003,Forbes2005}, and crystalline superfluid phases \cite{Petrov2007}. Quantum gas mixtures also serve as a precursor for the formation of ultracold heteronuclear molecules \cite{Ospelkaus2006,Papp2006,Ni2008,Weber2008,Klempt2008,Voigt2009}. Molecules of atoms with large mass ratios, such as rubidium and lithium, are predicted to have large dipole moments \cite{Aymar2005}, and may be useful for studies of dipolar quantum gases, precision measurements, and quantum computing.

These scientific opportunities motivate the development of techniques to produce a wide range of gas mixtures rapidly, robustly, and efficiently. Here, we focus on the first stage in producing such mixtures by laser cooling, namely the generation of slow atomic beams suitable for loading into magneto-optical traps.  Zeeman slowers \cite{Phillips1982} have  many advantages over competitive sources of single-element slow beams because they are compatible with a wide range of elements, including nonvolatile elements such as lithium that require specialized high-temperature ovens \cite{Beverini1989,*Lin1991,*Honda1999,*Chebakov2009,*McClelland2006,*Lu2009}; they are simple to operate and robust against drifts in laser alignment, polarization, power, and magnetic fields; they require only modest laser powers; and they produce high brightness and high flux beams.  The chief liability of a Zeeman slower is its initial design and construction and the possible geometric constraints of integrating it with the remainder of a cold-atom experiment. We mitigate these liabilities by building a single apparatus that produces simultaneous high brightness beams of two atomic elements with a large mass ratio. This paper demonstrates our success at producing continuous and overlapped slow beams of rubidium (\87Rb) and lithium (\7Li) using a specially designed Zeeman slower and two-elements oven.

\section{\label{sec:design}Design and Construction}
We summarize the operating principle and some design considerations of an increasing-field Zeeman slower to highlight the requirements for slowing multiple elements and to describe our strategy to satisfy these requirements \cite{Phillips1982,Barrett1991,Joffe1993}. Atoms in a Zeeman slower are decelerated and Doppler cooled by radiation pressure as they scatter photons out of a laser beam propagating counter to the atomic beam.  To maintain this deceleration and cooling with a constant laser frequency, the atomic beam is conducted through a spatially varying magnetic field that shifts the atomic resonance frequency by the Zeeman effect.  Atoms starting with longitudinal velocities below a design-determined capture velocity will be slowed continuously to a position-dependent velocity $v(x) \approx -(\Delta + \mu B(x)/\hbar)/k$ if the adiabatic radiation pressure acceleration
\begin{equation}
\aad = \mu\,\frac{d B}{d x}\frac{v}{\hbar k}
\end{equation}
is kept smaller than the spontaneous emission-limited acceleration, $\amax = v_r \Gamma/2$, a condition that depends on atomic properties including the magnetic moment $\mu$, the mass $m$, the recoil velocity $v_r = \hbar k/m$, the laser wavenumber $k=2\pi/\lambda$, and the spontaneous emission rate $\Gamma$, where $\Delta$ is the laser detuning from the zero-field atomic resonance, and $B(x)$ is the magnetic field at position $x$ (Tab.~\ref{tab:properties}) \footnote{Here we assume for simplicity that the Zeeman shift is linear, as is typical for the stretched state to which atoms are pumped by the slower laser.}. The adiabaticity requirement can be summarized with a dimensionless acceleration $\eta = \aad/\amax < 1$. Of these atomic properties, only the mass is substantially different between rubidium and lithum. The mass ratio implies that, for a given magnetic field and velocity, the lighter lithium can follow a much higher acceleration than rubidium with $\eta_\textrm{Rb}/\eta_\textrm{Li} = 17$. Such a large ratio must be considerd in the slower design to achieve high brightness of both elements.

\begin{table}[!htb]
\caption{\label{tab:properties}Atomic Properties and Typical Experimental Parameters}
\begin{ruledtabular}
\begin{tabular}{llcccc}
% \begin{tabular*}{\columnwidth}{@{\extracolsep{\fill}}ll@{\extracolsep{\fill}}l}\toprule
& Property & Symbol & Li & Rb & \\\hline
& D2 Wavelength        & $\lambda$        &  671    &  780  & nm    \\
& D2 Linewidth         & $\Gamma/{2\pi}$  &  5.9    &  6.1  & MHz   \\
& Atomic mass          & $m$              &  7.0    &  87   & amu   \\
& Recoil Velocity      & $v_r$            &  85     &  5.9  & mm/s  \\
& Maximum acceleration & $a_\textrm{max}$ &  1.6    & 0.11     & $10^6\;\textrm{m/s}^2$  \\
\hline
& Slower Laser Detuning         &&    1.1    &    0.9    & GHz    \\
& Slower Laser Power            &&    100    &    20     & mW    \\
% MOT Laser Detuning            &&    -7.6   &    -3.2   & $\Gamma$   \\
% MOT Repump Detuning           &&    -6.4   &    0      & $\Gamma$    \\
% MOT Laser Power               &&    120    &    120    & mW  \\
% MOT Gradient                  &&\multicolumn{2}{c}{12} & G/cm\\
\end{tabular}
\end{ruledtabular}
\end{table}

The atomic mass also has a strong effect on beam brightness. As atoms decelerate by scattering photons, the atomic velocity exhibits a random walk in the transverse velocity plane, with rms transverse velocity accumulation $v_\perp' \approx \sqrt{v_r v_\parallel/3}$, where $v_\parallel$ is the velocity of the atom upon entering the slower. This transverse heating leads to ``blooming'' or diverging of the atom beam upon exiting the slower into the angle $\theta\approx v_\perp'/v_\parallel'$, where $v_\parallel'$ is the final longitudinal velocity. Transverse heating is more severe for lithium than rubidium because of its lighter mass (higher $v_r$) and larger initial velocity \cite{Joffe1993}. The resultant blooming diminishes beam brightness, especially when the acceleration is far less than the maximum $\amaxli$.

% One implication of the lower mass and higher transverse heating of lithium is that a Zeeman slower designed for rubidium, with an acceleration near $\amaxrb$ and length sufficient to slow a significant fraction of the thermal distribution, would produce a low brightness beam of lithium. We note that in a lithium-only Zeeman slower, the higher transverse heating of lithium is offset by the shorter length and time of flight of the lithium beam.

% One implication of the lower mass and higher transverse heating of lithium is that a Zeeman slower designed solely for lithium would be far shorter than one designed solely for rubidium. For lithium, an increase in the maximum capture velocity achieved by lengthening the slower is outweighed by the consequent solid angle reduction and transverse heating at lengths on the order of 10 cm. For rubidium, the relatively low acceleration and low transverse heating dictate that the trade-off occur at a much longer length, on the order of 100 cm.

\begin{figure}[tb]
% \begin{center}
% \includegraphics[width=\figwidth\columnwidth]{theory}
\includegraphics{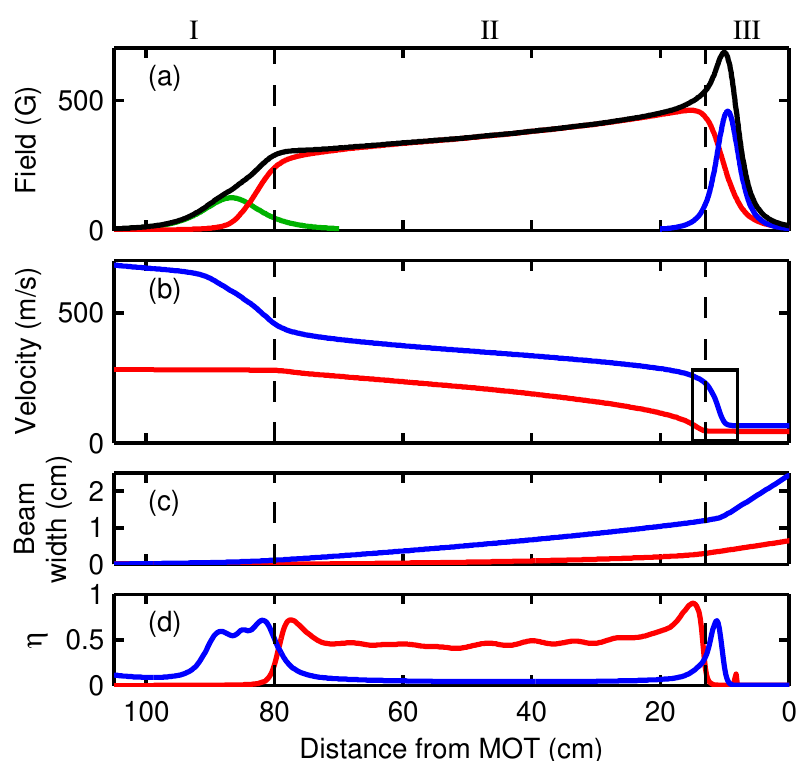}
% \end{center}
\caption{(color) (a) Magnetic field profile of the slower (black) generated by Stages I (green), II (red), and III (blue). Simulated trajectories of the (b) longitudinal velocity, (c) beam width, and (d) dimensionless acceleration $\eta=\aad/\amax$  of rubidium (red) and lithium (blue) atoms near the capture velocity of the Zeeman slower. The box in (b) emphasizes the key concept of our slower: the aggressive deceleration of lithium while rubidium disengages from the slower. Calculations assume zero initial beam width and transverse velocity.
% Inset: lithium atoms (blue) slow by $\Delta v_\textrm{III}$ in stage III while rubidium diabatically stops slowing.
\label{fig:theory}}
\end{figure}

Our Zeeman slower is designed and optimized to slow both rubidium and lithium simultaneously via an increasing-field three-stage design (Fig.~\ref{fig:theory}). The key concept of our design stems from the realization that the detrimental effects of transverse heating on the lighter lithium beam are most severe when the lithium beam is slow, requiring that the final stage of slowing the lithium beam be performed close to the maximum deceleration, at high magnetic field gradient, and in close proximity to the magneto-optical trap. In contrast, a faster lithium beam can be slowed without a great decrease in beam brightness at a lower deceleration and magnetic field gradient appropriate for the heavier element, rubidium. In other words, our strategy is to reduce the time of flight of lithium in a slower long enough to decelerate a significant fraction of the rubidium thermal distribution. Thus, we construct a Zeeman slower where a stage with moderate magnetic field gradient (Stage II), used to slow the rubidium beam to a low final velocity, is followed by a final high-gradient stage (Stage III) used for aggressive slowing of the lithium beam. At the transition between these stages, the rubidium beam is made to disengage from the slower by setting the field gradient and laser characteristics so that $\eta_\textrm{Rb} > 1 > \eta_\textrm{Li}$.  We note that such disengagement is common to decreasing-field Zeeman slowers \cite{Molenaar1997,Lison1999}. An additional stage (Stage I) is prepended to take advantage of the large, $\sim$300 gauss bias field required to separate the rubidium hyperfine levels \cite{Gunter2004} and to cool lithium entering into Stage II.

\begin{table*}[!t]
\caption{\label{tab:optimizaiton}Design Parameters}
\begin{ruledtabular}
\begin{tabular}{ccccccccc}
Stage & Target Atom & $\eta$ & Layers & ID (cm) & Wire size (in.) & Turns Per Inch & Current (A) & Resistance (m$\Omega$)\\\hline
I     & Li          & 0.78   & 3      & 12.7 &  1/8     & 3.5, 7 & 100 & 16 \\ %16 mOhms
II   & Rb          & 0.66   & 2, 14     & 2.5  & 1/8, 1/16 
                                                        & 3.5, 7, 14 & 6 & 4900 \\
III   & Li          & 0.68   & 2      & 2.5 & 1/8      & 3.5, 7 & 150 & 5.7 \\ %5.7mOhms
\end{tabular}
\end{ruledtabular}
\end{table*}
% grad comp = 4.3mOhms

We calculated the target magnetic field profile for our three-stage slower based on the desired length of each stage and a constant deceleration within the adiabaticity limit for the appropriate atom. The design parameters are listed in Table \ref{tab:optimizaiton}. The optimized winding was determined numerically by gradient descent on a cost function taking into account the deviation from the target field, the adiabaticity requirement, the rubidium diabaticity requirement in Stage III, and the peak field value. To produce the precise field demanded by such aggressive design parameters, the optimization allowed for 16 layers for Stage II and 6 layers for Stages I and III. The optimization method varied the current through each stage independently with fixed physical boundaries between each stage.

The construction of the desired winding pattern was done on a lathe tooled to guide wire at precise pitches, similar to Ref.~\onlinecite{Dedman2004}. Insulated 1/8 inch square hollow wire was used to provide high currents and water-cooling for Stages I and III, and the first two layers of Stage II. The remaining 14 layers of wire in the Stage II were wound from 6 independent lengths of Kapton-insulated solid 1/16 inch square wire. Stages II and III were placed over a 0.75 inch stainless steel vacuum tube.

The measured magnetic field of Stage II matches well with the calculated field for a perfectly wound coil. The measured rms deviation of the magnetic field is only 0.3\%, and the measured rms deviation of the gradient is 15\%, both dominated by measurement noise.

\section{\label{sec:setup}Experimental Setup}
%A cross-section of our experimental setup is shown in Fig.~\ref{fig:setup_sketch}. A dual-species oven produces overlapping collimated thermal atomic beams of \87Rb and \7Li. The atoms are decelerated in the Zeeman slower atoms and are then loaded into a double MOT.

The two-element beam source is an effusive oven with separate reservoirs for rubidium and lithium, inspired by Ref.~\onlinecite{Stan2005}. A single-reservoir design is impractical for the rubidium-lithium mixture owing to the vastly different vapor pressures of the two metals. At the typical lithium reservoir operating temperature, the equilibrium vapor pressure of rubidium is five orders of magnitude greater than that of lithium.  Instead, we use two reservoirs with independent temperature controls connected by a thin intermediate nozzle. The higher temperature lithium reservoir also serves as the mixing chamber. The rubidium reservoir, lithium reservoir, and intermediate nozzle are kept at $200^\circ\textrm{C}$, $400^\circ\textrm{C}$, and $450^\circ\textrm{C}$, respectively. Rubidium metal is introduced into its reservoir in a sealed glass ampoule that is broken after baking out the sealed vacuum chamber. Lithium metal is cleaned and then added directly to its reservoir. We seal the oven with annealed nickel gaskets because they resist corrosion by the hot lithium vapor.

\begin{figure*}[!t]
\includegraphics{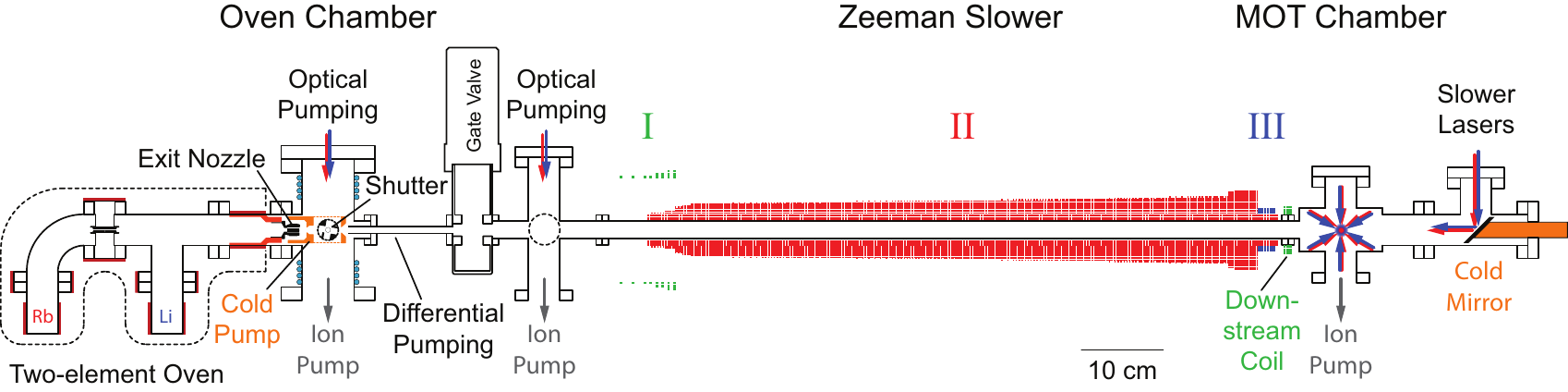} %[width=\textwidth]
\caption{(color) Experimental setup. See text for description. \label{fig:setup_sketch}}
\end{figure*}
We tested the operation of our two-element Zeeman slower in an experimental setup comprised of a two-element beam source, the slower, and a UHV region for magneto-optical trapping (Fig.~\ref{fig:setup_sketch}). The beam source is itself a rather specialized apparatus and so we provide more details on its design and performance. We then give a brief description of our vacuum chamber.

The two-element mixing chamber is followed by an oven nozzle comprised of a multichannel array of stainless steel tubes \cite{Pauly2000,Beijerinck1975}. A multichannel arrays is ideal for generating a large flux of a highly collimated beam because it has a large aspect ratio while maintaining a short length. The short length allows the oven to operate at a high pressure before collisions deteriorate the collimation. Our array is formed by aligning and then sintering $\sim$380  stainless steel tubes with inner diameter 160 \textmu m, outer diameter  310 \textmu m, and length 7 mm. A single tube with the same open area, aspect ratio, and Knudsen number would operate at 1/20 the pressure and be 140 mm long. We heat the nozzle to $450^\circ\textrm{C}$ to prevent clogging.

The nozzle directs collimated beams of rubidium and lithium through the oven chamber (Fig.~\ref{fig:setup_sketch}). A shutter can rotate to block the beams. A gate valve seals off the oven chamber for servicing and replenishing the reservoirs. Glass viewports before and after a differential pumping stage provide optical access, allowing for transverse optical pumping of the atomic beams. After the oven chamber, the beams are slowed longitudinally in the Zeeman slower and captured in a magneto-optical trapping UHV chamber (MOT chamber). Atoms not captured stick to a cold in-vacuum gold mirror that reflects the Zeeman slower laser onto the atomic beam axis. The mirror allows the laser light to enter the vacuum chamber through a glass viewport that is not directly exposed to the atomic beam, preventing its occlusion and corrosion.

The vacuum system maintains a pressure differential between the oven chamber at $10^{-9}\;\textrm{mbar}$ and the MOT chamber at $10^{-11}\;\textrm{mbar}$. The pressure in both chambers is dominated by hydrogen. The oven chamber is pumped by a $-20\:{}^\circ\textrm{C}$ cold plate, which efficiently pumps rubidium and lithium, and an ion pump protected from the alkali vapor by a water-cooled chevron baffle. Differential pumping between the oven and MOT chamber is maintained through a thin tube with a hydrogen conductance of $1.3\;\textrm{L/s}$ and an intermediate ion pump. The Zeeman slower itself has a hydrogen conductance of $3\;\textrm{L/s}$ and provides further differential pumping. The MOT chamber is pumped by an ion pump, titanium sublimation pump, and non-evaporable getter. The in-vacuum mirror in the MOT chamber is cooled to $-20\:{}^\circ\textrm{C}$ and pumps rubidium and lithium.

% Differential tube is 5 inches long with an ID of 0.28", F = 1.3 L/s
% Zeeman slower is 35" long with an ID=0.72", F = 3.2 L/s

\section{\label{sec:results}Results and Discussion}
We characterize our Zeeman slower by measuring the flux of rubidium and lithium with two methods. First, we measure the velocity distribution by illuminating the atomic beams in the MOT chamber with probe light at $45^\circ$ and collecting fluorescence with a photomultiplier tube, with sensitivity enhanced by lock-in detection. Second, we measure the MOT loading rate by collecting fluoresced trapping light.

\begin{figure}[!htb]
% \begin{center}
\includegraphics{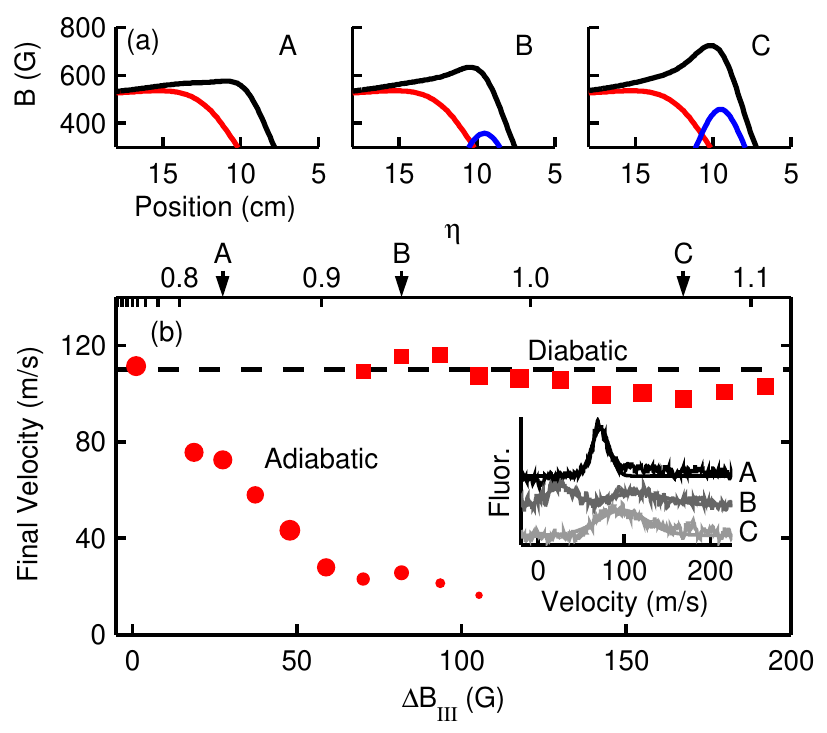} %[width=\figwidth\columnwidth]
% \end{center}
\caption{(color) Diabatic transition of the rubidium beam from the Zeeman slower. (a) Magnetic field profiles generated by Stage II (red), Stage III (blue), and the total field (black). Position is measured from the MOT. (b) Final velocity of the adiabatic (circles) and diabatic (squares) beams. The bottom axis is the magnetic field ramp generated in Stage III. The top axis is the dimensionless peak acceleration $\eta=\aad/\amax$. Marker areas are proportional to the beam flux. Inset: velocity distributions from fluorescent scattering for \textsf{A} ($\eta <0.9$), \textsf{B} ($0.9 < \eta < 1.0$), and \textsf{C} ($1.0 < \eta$). Traces are offset for clarity.\label{fig:breakthrough}}
\end{figure}

We first test the key concept of our slower design, the disengagement of the rubidium beam from the Zeeman slower due to a violation of the adiabaticity requirement $\eta < 1$ (Fig.~\ref{fig:theory} (b), inset). The magnetic field generated by Stage II and laser detuning sets a rubidium beam of $110\;\textrm{m/s}$ at the end of Stage II.  We gradually increase the magnitude of the peak magnetic field by $\Delta B_\textrm{III}$ added by the windings of Stage III, and monitor the Doppler distribution of the rubidium beam that emerges at the end of the slower into the MOT region (Fig.~\ref{fig:breakthrough}). For small $\Delta B_\textrm{III}$, the rubidium beam follows the Zeeman slower field adiabatically, as $\eta$ remains below unity, leading to a gradual reduction in the beam's final velocity. For intermediate $\Delta B_\textrm{III}$, with $0.9 < \eta < 1.0$, the beam bifurcates into two resolved velocity classes, with one portion adiabatically following the slower field to ever lower velocities while the other portion remains at the velocity at which it enters Stage III, having now diabatically disengaged from the slowing process.  For large $\Delta B_\textrm{III}$, with $\eta > 1$, the diabatic transition is complete as the entire beam remains at its original velocity.  This beam is therefore unaffected by the high-gradient magnetic field of Stage III which can now be employed to slow the lithium beam selectively.

\begin{figure}[!htb]
% \begin{center}
\includegraphics{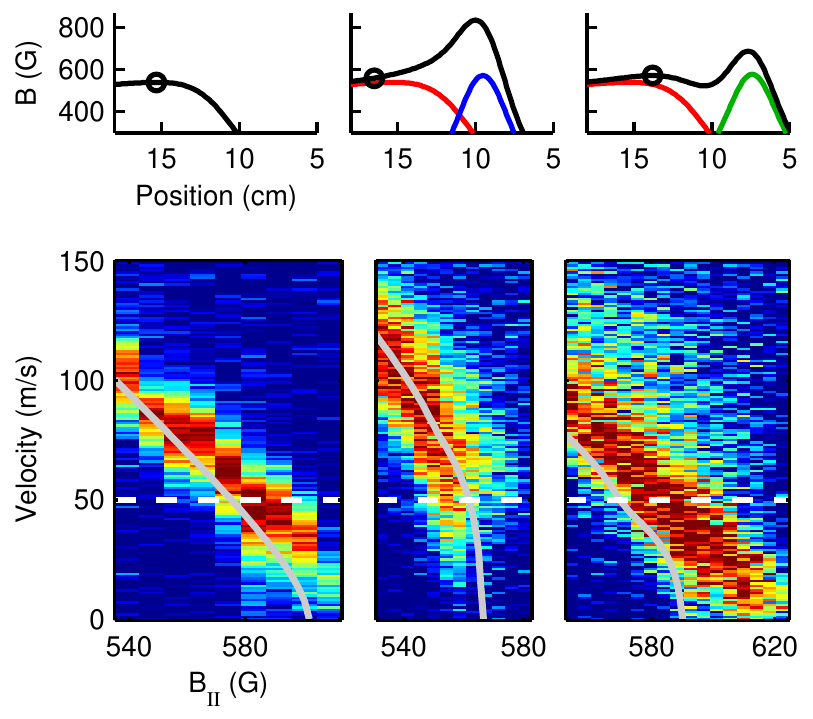}
% \end{center}
\caption{A comparison of three operating modes of the slower: a rubidium-only mode (left), a two-element mode (center), and a two-element mode with an alternative coil configuration (right). Top: Magnetic field profiles generated by Stage II (red), Stage III (blue), an additional downstream coil (green), and the total field (black). Position is measured from the MOT. Circles indicate the point at which rubidium ceases to slow at a field of $B_\textrm{II}$ at the end of Stage II. Bottom: Corresponding rubidium velocity distributions (vertical axis) for different initial velocities set by $B_\textrm{II}$. Solid gray lines are theory and dashed white lines are the estimated MOT capture velocity. \label{fig:rb_velocitymap}}
\end{figure}

\begin{figure}[!htb]
% \begin{center}
\includegraphics{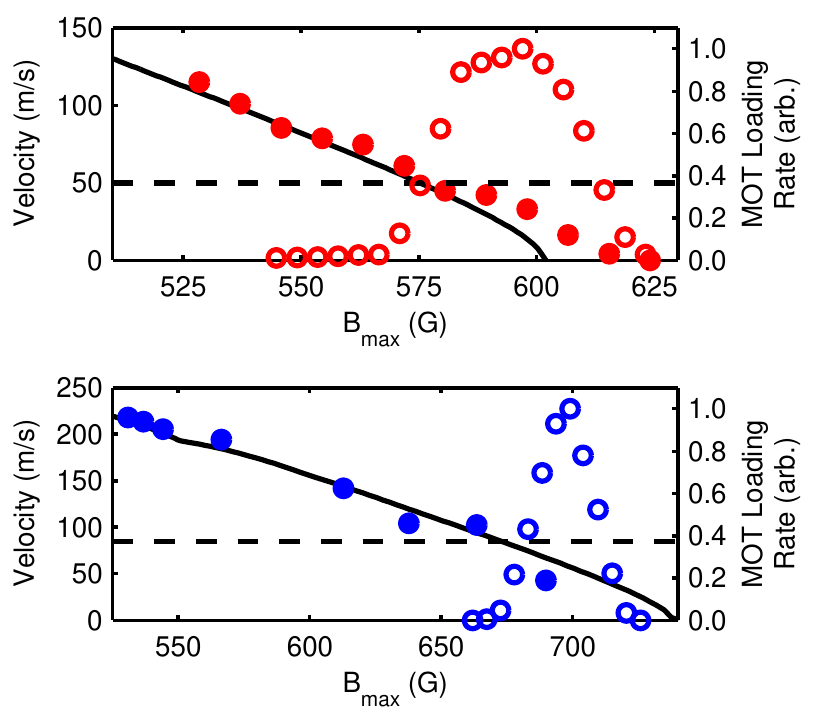} %[width=\figwidth\columnwidth]
% \end{center}
\caption{Final velocity (filled circles, left axis) and MOT loading rates (open circles, right axis) of rubidium (top, red) and lithium (bottom, blue). The black solid lines are calculations of the final velocity with no adjustable parameters. The dashed horizontal lines are estimates of the MOT capture velocity.
\label{fig:rbli_velocity}}
\end{figure}

While the diabatic transition of the rubidium beam is successful at high final velocities,  this transition is accompanied by a significant depletion at low final velocities.  This depletion is illustrated in Fig.~\ref{fig:rb_velocitymap}, where we present the Doppler distributions of the rubidium beam observed with constant $\Delta B_\textrm{III}$ but variable velocities at the end of Stage II, set by the final magnetic field $B_\textrm{II}$.  With $\Delta B_\textrm{III} = 0$, the slower is operated as a conventional single-element Zeeman slower (Fig.~\ref{fig:rb_velocitymap}, left).  The final beam velocity can be smoothly tuned by varying $B_\textrm{II}$.  We have used such a beam to load a rubidium-only magneto-optical trap (Fig.~\ref{fig:rbli_velocity}), finding a MOT capture velocity of about 50 m/s and confirming a numerical model that predicts both the final velocity of the slowed beam and the capture velocity of the MOT.

In contrast, using a value for $\Delta B_\textrm{III}$ compatible with slowing lithium in Stage III, a large fraction of the rubidium beam flux is lost at low initial velocities (Fig.~\ref{fig:rb_velocitymap}, center).  We hypothesize that we are limited by the magnetic field curvature $B''$ provided by the Stage III coils. A rubidium atom spends a time
\begin{equation}
\tau \simeq \frac{m \Gamma (\eta_2-\eta_1)}{2 \mu B''}
\frac{v_r^2}{v_\textrm{II}^2}
\end{equation}
in the continuous transition from the adiabatic $1 > \eta_1$ to the diabatic $\eta_2 > 1$. During this time, the atom experiences ``laser heating'' because it is blue-detuned from and near-resonant to the laser. A successful disengagement requires that this time be short. A fixed value of $B''$ therefore sets a lower limit on the velocity $v_\textrm{II}$ at which the rubidium beam may disengage from the slower and emerge at the MOT region.

To test this idea, we implemented the alternative slower configuration shown on the right of Fig.~\ref{fig:rb_velocitymap}.  Replacing the Stage III coil with a downstream coil, the magnetic field diminishes at the end of Stage II, allowing the slowed rubidium beam to continue along the slower to a point where the laser light is brought again rapidly across the atomic resonance at a high field gradient, with $\eta > 1$. We observe a slow rubidium beam well below the MOT capture velocity. We note that this field configuration has also been used to avoid heating and increase the flux of low-velocity atoms in decreasing-field slowers \cite{Molenaar1997,Lison1999}. However, this particular field configuration is not compatible with slowing lithium and not used in our setup.

\begin{figure}[!b]
% \begin{center} 
\includegraphics{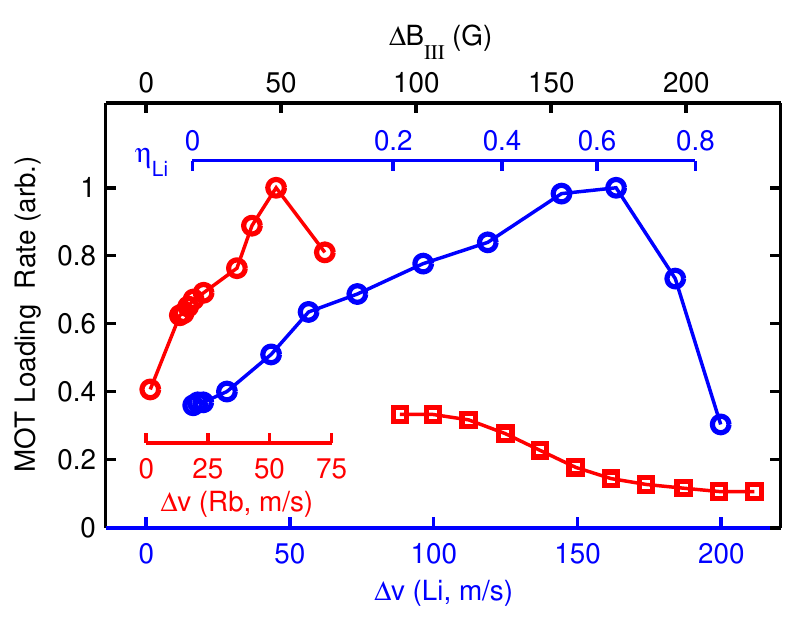} %[width=\figwidth\columnwidth]
% \end{center}
\caption{MOT loading rates of lithium (red circles) and rubidium (blue circles and squares) as a function of the field ramp $\Delta B_\textrm{III}$ (top axis), velocity decrease $\Delta v$ (blue and red bottom axes for rubidium and lithium, respectively), and lithium peak deceleration $\eta_\textrm{Li}$ (top red axis) in Stage III. The final velocity is kept constant for each point by adjusting the Stage II field. Stage III slows rubidium atoms when the field is sufficiently small (blue circles). Above 70\;G rubidium is not slowed in Stage III (blue squares). Lines are guide to the eye.\label{fig:boost_loadingrate}}
\end{figure}

Substantial slowing of lithium in Stage III enhances the lithium MOT loading rate (Fig.~\ref{fig:boost_loadingrate}). We measure the MOT loading rates of lithium and rubidium as a function of $\Delta B_\textrm{III}$, which sets the velocity decrease of the final stage. The maximum MOT loading rate of lithium occurs for a substantial $\Delta v_\textrm{III} \approx 150\;\textrm{m/s}$ decrease of the final velocity in Stage III, which corresponds to an $\eta_\textrm{Li} \approx 0.6$. We note that the adiabatic rubidium beam also shows enhanced loading for small $\Delta B_\textrm{III}$.

% \begin{figure}[!htb]
% % \begin{center}
% \includegraphics[width=\figwidth\columnwidth]{motloading}
% % \end{center}
% \caption{Rubidium (blue) and Lithium (red) MOT loading traces. Simultaneous MOTs exhibit interspecies losses.
% \label{fig:motloading}}
% \end{figure}

With two-element slow beams we can load a double-MOT. For single-element operation we measure a maximum MOT loading rate of $3\times 10^8\;\mathrm{s}^{-1}$ rubidium atoms or $4\times 10^7\;\mathrm{s}^{-1}$ lithium atoms, which yield a MOT saturated at $3\times 10^9$ rubidium atoms or $8\times 10^8$ lithium atoms. At the peak lithium flux, with a diabatic disengagement of rubidium, we load $15-30\%$ of the peak rubidium flux. The two MOTs exhibit enhanced interspecies loss because of inelastic rubidium-lithium collisions. The losses can be reduced by loading rubidium for a short time and offsetting MOT centers with imbalanced cooling lasers. In the future we may utilize a dark SPOT MOT \cite{Ketterle1993} to reduce losses by decreasing the density of excited rubidium atoms.

% \newpage
\section{\label{sec:conclusion}Conclusion}
We have demonstrated a two-element oven and Zeeman slower that produce simultaneous and overlapped slow beams of rubidium and lithium. The key concept is a high-gradient final stage (Stage III) that aggressively decelerates lithium after a low-gradient stage (Stage II) appropriate for decelerating rubidium. The transition between the stages should be adiabatic for lithium but diabatic for rubidium so that rubidium is unaffected by the high-gradient stage. Our implementation produces a depleted low-velocity rubidium beam because of limited magnetic field curvature during the transition. However, we have shown that an improved design with a dip in the magnetic field successfully acheives the selective diabatic transition of rubidium at velocities that can be captured in a MOT.

\section{\label{sec:acknowledgements}Acknowledgements}
We thank J. Daniels, J. Guzman, and X. Wu for assistance in the early stages of this work. This work was supported by DARPA (Grant No.\ 49467-PHDRP) and DTRA (Contract No.\ HDTRA1-09-1-0020). GEM acknowledges fellowship support from the Hertz Foundation. A\"O acknowledges support from Deutsche Forschungsgemeinschaft. EV acknowledges support from Bundesministerium f\"ur Bildung und Forschung. DMSK acknowledges support from the Miller Institute for Basic Research in Science. 

% for assistance in 

% \bibliographystyle{aipnum4-1}
% \bibliographystyle{apsrev4-1}
\bibliographystyle{mypra}
% \nocite{*}
\bibliography{slower}

\newpage

\end{document}